\documentclass[conference]{acmsiggraph}

\usepackage{stfloats}
 
\title{4Doodle: Two-handed Gestures for Immersive Sketching 
 of Architectural Models}

\author{
Fernando Fonseca\thanks{e-mail:fonsecafja@gmail.com}
\and
Maurício Sousa\thanks{e-mail:mauricio.sousa@utoronto.ca}
\and
Daniel Mendes\thanks{email:danielmendes@fe.up.pt}
\and
Alfredo Ferreira\thanks{email:alfredo.ferreira@tecnico.ulisboa.pt}
\and
Joaquim Jorge\thanks{e-mail:jaj@inesc-id.pt}\\INESC-ID T\'{e}cnico Lisboa, Av Rovisco Pais S/N 1049-001 Lisboa Portugal
}

\pdfauthor{anonymous}


\begin{document}


\maketitle

\begin{abstract}

Three-dimensional immersive sketching for content creation and modeling has been studied for some time. However, research in this domain mainly focused on CAVE-like scenarios. These setups can be expensive and offer a narrow interaction space. Building more affordable setups using head-mounted displays is possible, allowing greater immersion and a larger space for user physical movements. This paper presents a fully immersive environment using bi-manual gestures to sketch and create content freely in the virtual world. This approach can be applied to many scenarios, allowing people to express their ideas or review existing designs. To cope with known motor difficulties and inaccuracy of freehand 3D sketching, we explore proxy geometry and a laser-like metaphor to draw content directly from models and create content surfaces. Our current prototype offers 24 cubic meters for movement, limited by the room size. It features infinite virtual drawing space through pan and scale techniques and is larger than the typical 6-sided cave at a fraction of the cost. In a preliminary study conducted with architects and engineers, our system showed a clear promise as a tool for sketching and 3D content creation in virtual reality with a great emphasis on bi-manual gestures.



\end{abstract}

\begin{CRcatlist}
    \CRcat{H.5.2}{Information Interfaces and Presentation}{User Interfaces}{Input devices and strategies}
    \CRcat{H.5.2}{Information Interfaces and Presentation}{User Interfaces}{Interaction styles}
    \CRcat{I.3.4}{Computer Graphics}{Graphics Utilities}{Paint systems}
    \CRcat{I.3.7}{Computer Graphics}{Three-Dimensional Graphics and Realism}{Virtual Reality}
\end{CRcatlist}

\keywordlist


\TOGlinkslist


\copyrightspace

\section{Introduction}


Freehand sketching on paper is a highly trained sensorimotor skill used daily by most designers and architects to provide an understandable view of their ideas. It appears in different project phases serving the cognitive processes~\cite{Lim:2003}. For example, in some architectural work-flows, sketches are used for creative purposes, materializing initial ideas and concepts that will eventually lead to complete buildings, similar to the known works of Frank Gehry\footnote{Gehry Partners, LLP: \url{https://www.foga.com/}}. Moreover, sketches can be used when reviewing existing projects in architecture, engineering, and other fields, either complementing the model with new shapes or creating annotations of important remarks.

Several solutions for 2D sketching exist, following the pen and paper metaphor~\cite{olsen2009sketch}. On the other hand, the interest from both academic research and industry for extending sketches to the third dimension has been increasing~\cite{Deisinger:2000,Keefe:2001,Keefe:2008}. While a skill such as pen and paper sketching that has been mastered since the early stages of our lives can hardly be compared with a mostly unfamiliar combination of virtual environments and 3D freehand sketching techniques, these later approaches allow for a more direct 3D content creation, allowing the user to directly sketch in 3D and, at the same time, visualize the virtual environment from several angles.

While desktop modeling solutions have been a standard until now, they offer a very small space for visualizing and interacting with digital content. Also, the two-dimensional view of the virtual environment provides limited perspective, offering no depth of perception. Stereoscopic tabletops can overcome some challenges but still have a small interaction space and only allow for scale model interaction metaphors. Other solutions, such as CAVE systems or head-mounted displays, can co-locate the user and the model at a one-to-one scale. These immersive solutions allow users to use their movements to navigate the virtual environment and sketch in first-person, thus creating a more intimate relationship between the creator and the artifact.

\begin{figure}[t!]
  \centering
    \includegraphics[width=\columnwidth]{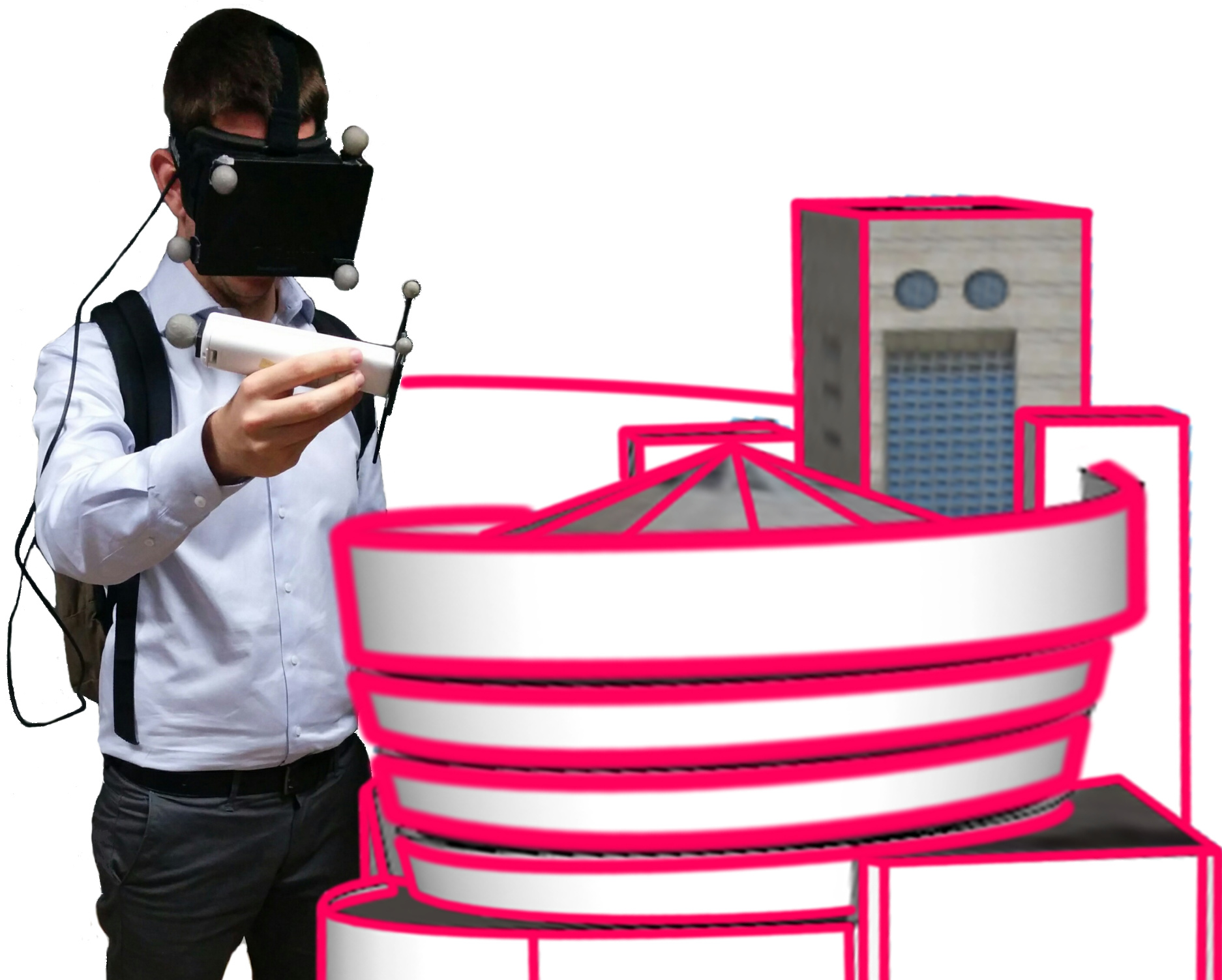}
 \caption{Our vision of the immersive solution for sketching on 3D content.}
    \label{fig:master_vision}
\end{figure}


Previous works for 3D sketching mainly rely on tracked tools within CAVE-like scenarios~\cite{Israel:2009,Keefe:2001,Keefe:2008,Israel:2013}, but assembling an entire CAVE setup can be expensive. This paper presents an affordable system that offers a whole room for freehand 3D sketching. We adopt the Oculus Rift\footnote{\url{http://www.oculusvr.com}} head-mounted display and an OptiTrack\footnote{\url{http://www.naturalpoint.com/optitrack}} more system to create a prototype that is completely immersive and accurate, while achieving a great relation between cost and physical interaction volume. Indeed, the built-in accelerometer/gyro system already contained in the HMD, while great for simple locomotion and exploration tasks, is unsuitable for precision co-location due to drift and stability. The optical tracking system provides fast and accurate locations of different objects in 3-space.

We use the head-mounted display within a tracking space of 24 cubic meters, providing the ability to walk and rotate inside the virtual world. Combining this with a set of devices enables free bi-manual 3D sketching inside that virtual world, creating a powerful tool with countless different applications, as depicted in Figure~\ref{fig:master_vision}. We can engage users in various scenarios by changing the system's background and virtual models, creating different possibilities. For instance, our approach supports quickly materializing new ideas or reviewing existing models, enabling users to directly sketch on existing model surfaces in an immersive 3D design review system, which previous work has already shown to be useful in the design review process~\cite{improve:2007}.

In our prototype, the user can freely navigate and sketch in the virtual space or the existing virtual surfaces. However, the lack of drawing surfaces makes it difficult to place content in three-dimensional spaces precisely. To ease this task, we created a proxy geometry tool similar to the one presented in 3D6B Editor~\cite{Kallio:2005} and a point laser-like tool to draw in distant surfaces. We also developed bi-manual gestures to allow the creation of fast objects and primitives to help express ideas with volume. To enable design review thought navigation in a one-to-one scale, we can control the virtual world using pan and scale actions, similar to Grabbing the Air~\cite{Bowman:2004,Stoakley:1995}. With these, our system features 24 cubic meters for physical movement, limited by the room size, and infinite virtual sketching space.

\section{Related Work}

Addressing the challenge of sketching 3D in a 2D space, Kallio et al.~\cite{Kallio:2005} developed a system that enables creating sketches with complex non-planar 3D strokes while retaining the essence of a pen and paper-based sketching projecting 2D input form a single viewport to a grid surface that can be manipulated in real-time. Conntero et al.~\cite{CIGRO} showed that sketching polyhedra with a single view reduced instruction set user interface is possible. Other studies showed that freehand drawing in the air needs additional motor coordination compared to pen and paper counterparts~\cite {Deisinger:2000,Makela2:2004,Makela:2004}. These air-sketching studies were also revealed to be less detailed and accurate. More recently, Wiese et al.~\cite{Wiese:2010} conducted a study investigating the learnability in immersive freehand sketching; their results suggest significant sketch quality enhancement over time. However, no change in the time needed to complete the given tasks was noticed.
 
With CavePainting, Keefe et al.~\cite{Keefe:2001} showed that creating artistic content in a virtual reality environment is possible by using only a collection of simple brush tools. Makela et al.~\cite{Makela:2004} also stated that their study subjects affirmed that this immersive sketching system offers new potential for artistic expression.
Moreover, a study conducted by Israel et al.~\cite{Israel:2009} showed that three-dimensional line-based sketching, using pen interactions to keep the similarities of 2D sketching on paper,  provided results regarding the benefits of 3D sketching and its special features in the design process.
Also, Israel et al.~\cite{Israel:2013} allowed designers to model life-size forms in real-time with four different approaches, searching the potential of immersive modeling even though no materiality is present. The results suggest that it is possible to achieve professional manipulative skills within virtual environments comparable to their work with physical material.
Perkunder et al.~\cite{Perkunder:2010} also used 3D sketched strokes as input in an immersive 3D environment to support automatic shape creations. They compared line-based sketching and sketch-based modeling, concluding that the usability of a creative sketching task was perceived to be higher for line-based sketching than for sketch-based modeling.

Combining augmented reality through head-mounted displays with cameras and freeform fabrication technology, Yee et al.~\cite{Yee:2009} allowed 3D sketching with live 3D references. The authors used the Optitrack motion capture system to track the user's head and wand position and orientation. Cabral et al.~\cite{Cabral:2011} developed an immersive system that allows the manipulation of simple house elements, such as rooms, doors, and windows. This work showed that immersive systems can also be used for the initial conceptual design of architectural models. 

Regarding selection and manipulation techniques for 3D interactions, some methods were developed using virtual pointer metaphor-based techniques. Bowman et al.~\cite{Bowman:1997} presented a ray-casting technique where the user can select an object by pointing a ray to it. Another technique to select object is based in flashlight and aperture, in this method user uses a pointing ray as well but it replaces the virtual ray with a conic selection volume. A modification of this technique originated the aperture technique that allows the user to control the spread of the selection volume, and the pointing directions are defined by the 3D position of the tracked head location and the position of the hand sensor~\cite{Forsberg:1996}.
Another selection technique consists of the image-plane family~\cite{Pierce:1997}, allowing the user to select an object using only 2DOF by selecting and manipulating the object in the projection image.

Moreover, virtual hand techniques were developed to allow direct manipulation. A simple virtual hand technique simulates everyday object interactions, making them natural. This technique directly maps the user's hand motion to a virtual hand's motion in a virtual environment~\cite{Bowman:2004}. Poupyrev et al. attempted to improve this technique and developed Go-Go~\cite{Poupyrev:1996}, allowing users to change the virtual arm's length interactively.

Regarding travel task scenarios using manual manipulation techniques, one explored method is to allow the user to grab the air. In this technique, when the user gestures to grab at any point in the virtual world and then moves their hand, the world as a whole moves while the viewpoint remains stationary~\cite{Bowman:2004}.

Other methods to control the user viewpoint have been explored, and the Head Tracking method to change the viewport position and orientation is a good example of this~\cite{Bakker:1998}. This approach is the most natural and direct way to change the user viewpoint, and it has been shown that physical turning leads to higher levels of spatial orientation than virtual turning.
Our system combines most of the techniques discussed to achieve an easy-to-use immersive sketching system.

\section{Setup Overview}

\begin{figure}[t!]
  \centering
  \includegraphics[width=\linewidth]{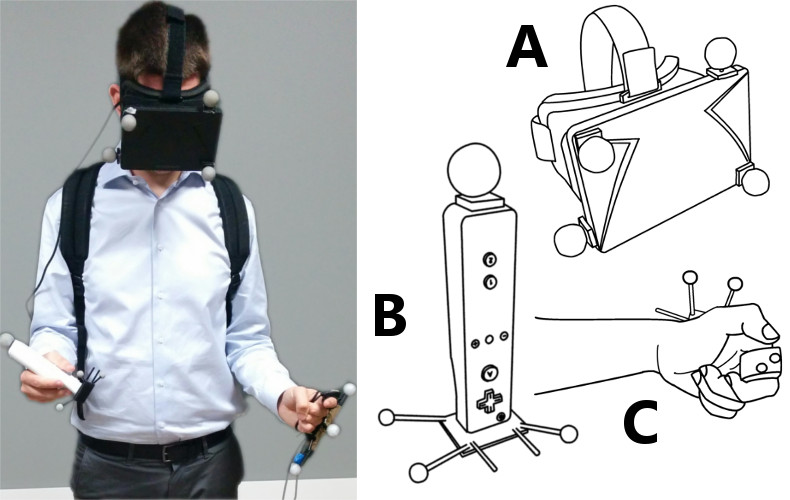}
  \caption{Our setup consists of an Oculus Rift (A), a Wiimote (B), and a ring wireless mouse (C). All these devices are tracked using markers.}
  \label{fig:setup}
\end{figure}

Our vision is to allow the user to perform freehand sketching inside a virtual space, taking advantage of an entire physical room. We use OptiTrack, a motion-tracking system that allows rigid-body triangulation and provides real-time position and orientation to follow the user inside the room.
Our system uses ten cameras strategically positioned around the room, covering roughly 24 cubic meters (4 x 3 x 2 meters) of interaction space. The physical dimensions and encumbrance of the room used limit this setup. If a larger room were available, our setup could support larger design spaces, limited only by camera resolution.

The system has Oculus Rift, a head-mounted display solution that offers stereoscopic visualization out of the box to provide an immersive visualization of the virtual space.

Combining the head-mounted display with visual markers that can be tracked by our motion capture system, as depicted in Figure~\ref{fig:setup}.A, the system can map the users' head position and orientation from the physical to the virtual world.
This ability, allied with a small battery and a wireless HMDI device inside a small backpack, frees the user of any wire limitation, turning the Oculus Rift and our system wireless with an insignificant input lag. With the immersive stereoscopic visualization provided by this head-mounted display, we can allow the user to move on to the physical world and perceive the same movement in the virtual space.

Visual markers were fitted in a Wiimote to mimic a common pen, as depicted in Figure~\ref{fig:setup}.B. 
This combination allows the physical device to be represented in the virtual world, matching its movements and orientation in space. Using this pen in the dominant hand, the user has haptic feedback to freehand sketch in the virtual world.

In order to add more features to our prototype, the user's non-dominant hand is instrumented with visual markers and two buttons, as outlined in Figure~\ref{fig:setup}.C. This technique opens more input possibilities to the system and enables non-dominant hand gestures.
Since both hands have markers, our system also explores bi-manual gestures.

Our prototype combines the data received from the OptiTrack system, the signals from both-hand buttons, and the Oculus Rift in the Unity3D rendering engine.

\section{4Doodle}

\begin{figure}[b!]
  \centering
  \includegraphics[width=\linewidth]{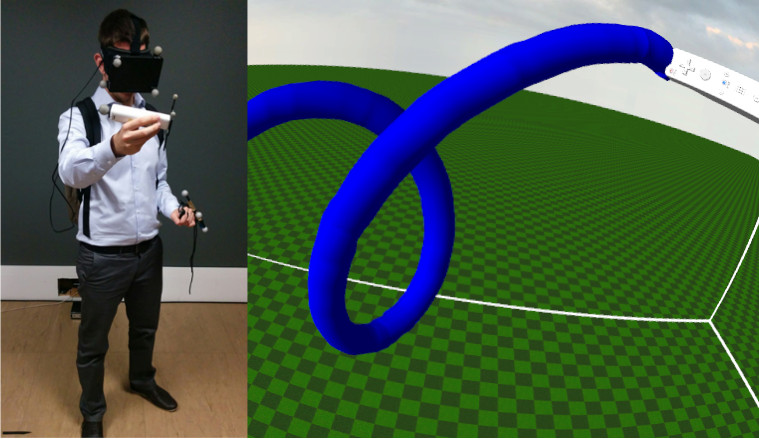}
  \caption{user sketching a line in our system. A user is on the left, and his virtual point of view is on the right.}
  \label{fig:screen}
\end{figure}

The developed system, depicted in Figure~\ref{fig:screen}, followed an iterative process, with several tests to obtain feedback and improve the system. Some of the process thoughts and decisions will be demonstrated in this section.
Since the user is immersed in the virtual world and does not see his physical surroundings, we provide some virtual visual clues.
The most important is the drawing pen, which is physically located in the user's dominant hand and works as a drawing instrument that mimics all the user's hand movements and rotations. The virtual representation is the same as the real device, the Wiimote 3D model.

When the user touches the Wiimote buttons, they change color to give the user additional visual feedback. Although there are many buttons, we decided to use only two so the user's hand doesn't need to constantly change position while searching for other buttons on the remote controller and can fully concentrate on creating virtual content.

On the virtual Wiimote representation's front edge is a small sphere representing the interaction point. This sphere changes in size and color depending on the chosen writing trace size, color, and interaction mode. 

Our system enables different interaction modes, including a freehand sketching mode in which the user can freely draw in three-dimensional space.
Another mode is laser sketching, which differs from the normal mode since the sketch is projected directly on the surface where the pen is pointing. 
The prototype also has a mode to create primitives and a mode to allow the user to select and manipulate any virtual 3D object pre-existent or made in the virtual scenario. These interactions will be further explained in the next sections.

\subsection{Air Sketching}

Our initial approach for sketching lines was cylindrical billboards. This solution worked well when the user was static creating his sketch, although when the user started to move around, the billboard effect began to notice a strange feeling that the sketch was moving along with the user. This sensation is caused by the billboard's properties of rotating themselves to always face the camera and the fact that we have two cameras rendering at the same time, one for each eye. 
To solve this problem, we adopted a tube approach, creating cylinders instead of billboard lines. This gives the sketch a sense of tridimensional volume that was not very clear with the billboard approach.
We also used a self-illuminated shader, so shadows did not affect the sketches.

\begin{figure*}[ht!]
  \centering
  \includegraphics[width=\linewidth,trim = 0mm 5mm 00mm 10mm, clip]{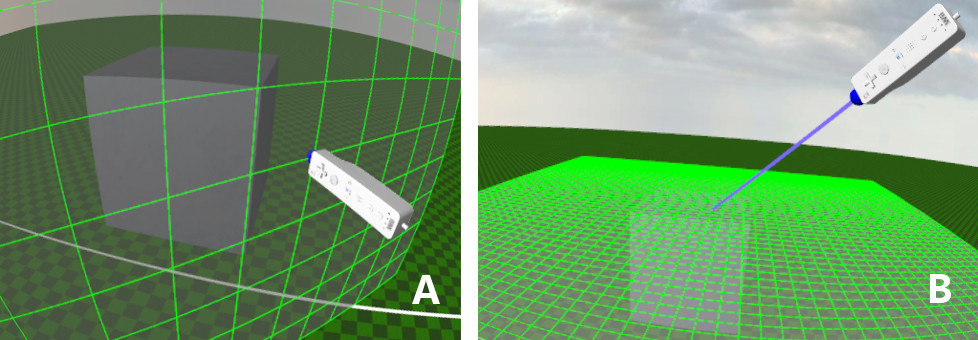}
  \caption{Two possibilities to place the proxy geometry plane: directly positioning it with the pen (A) or pointing at the desired surface (B).}
  \label{fig:plane}
\end{figure*}

Sketching in a 3D space instead of a 2D paper is more difficult~\cite{Deisinger:2000,Makela2:2004,Makela:2004}. Aware of this, we developed a tool to add restrictions and help with more accurate tasks. Inspired by the work in the 3D6B editor~\cite{Kallio:2005}, we created a plane to help restrict the drawing to a 2D space.
This tool allows the user to position a semi-transparent grid plane anywhere in the virtual space and sketch it restricted to that plane surface.
Initially, we allowed a complete orientation-free plane, where the user could position the plane at any angle in the three axes. However, by observation and commentaries, we noticed that users usually use this tool to help sketch objects and when they desire to accomplish perpendicular or parallel lines. With this in mind, we restrict the plane tool's orientation to snap at 15 degrees on each axis. This constraint allows easy perpendicular or parallel restrictions and better feedback from the subjects. This tool is depicted in the Figure~\ref{fig:plane}.A.

Since our system aims to provide the ability to make 3D sketches on existing models, we developed another method to use the plane tool. Freely positioning a plane exactly on a surface is as difficult as sketching in 3D. To ease this task, we give the possibility to position the semi-transparent grid plane directly in any existing surface in the virtual world and automatically adjust the grid size to the surface size.
To position the plane, the user uses the pen as a virtual pointer, aiming at the desired surface (depicted in Figure~\ref{fig:plane}.B). This approach allows users to sketch on top of any object's face or extend the drawing along that infinite virtual plane created by our tool.
These options are accessible in the two buttons on the user's non-dominant hand while the plane is toggled on.
After the plane is positioned, to give visual feedback that the pen is snapped to the plane's surface, an arrow is displayed on the back end of the pen, giving the user direct feedback. 
The user can then sketch in two different modes, using the distance between the hands to switch between them. If the hands are closer than a threshold, the pen will draw straight lines and snap to the grid squares on the auxiliary plane tool. Otherwise, the pen will freely draw in the 2D plane's space if the hands are farther away than a specified threshold.

\begin{figure}[b!]
  \centering
  \includegraphics[width=\linewidth,trim = 0mm 30mm 00mm 28mm, clip]{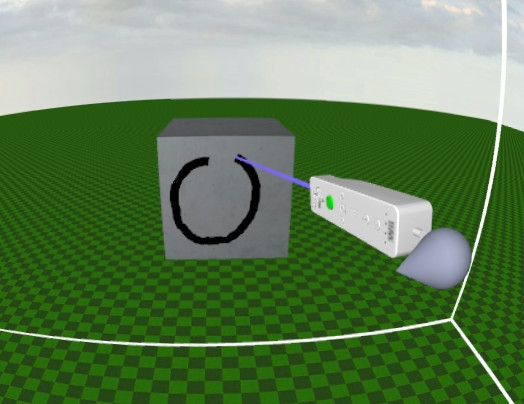}
  \caption{Sketching on existing 3D model's surfaces using the laser tool.}
  \label{fig:laser}
\end{figure}

Using a laser-like metaphor, we give users another way to draw on the desired surfaces. In this mode, an auxiliary ray is shown to give the user visual feedback on where they will draw. The user can sketch directly on any surface without restrictions, Figure~\ref{fig:laser}.
When the user starts to sketch in these modes, an ink drop object is displayed in the back end of the virtual Wiimote to give additional visual feedback.

\subsection{Painter's Palette}

We can create a virtual dynamic menu attached to the users' hands by equipping the user's non-dominant hand with a rigid body marker. The menu attached to the hand is displayed by facing the palm upwards, as shown in Figure~\ref{fig:screen}; facing the palm downwards is hidden. 
Using this implementation of a painter's palette metaphor, we can keep the field of view clear when the menu is not needed and avoid unwanted menu selections while sketching.

Our menu gives a set of different options to change modes and tools. 
The options allow you to change the sketching line size and color, toggle the auxiliary grid plane display, control the world, create primitives, switch between laser and free sketching, or select and manipulate objects.
The user hovers the button with the pen's edge to select any of these options. A highlight will appear, giving visual feedback and helping avoid unwanted selections. The user must press the sketching button on his dominant hand to confirm the choice.
Also, to avoid starting to draw while trying to select an option, we deactivate the ability to sketch while the pen is floating near the menu.
If the color option is selected, a sub-menu will appear, which will allow you to change the sketching color.
The currently selected color appears in the center of the menu, as well as the Wiimote tip. 
The lines size option also pops a sub-menu that offers different line sizes with toggle buttons.

\subsection{Grabbing the air}

\begin{figure*}[ht!]
  \centering
  \includegraphics[width=\linewidth,trim = 0mm 0mm 00mm 0mm, clip]{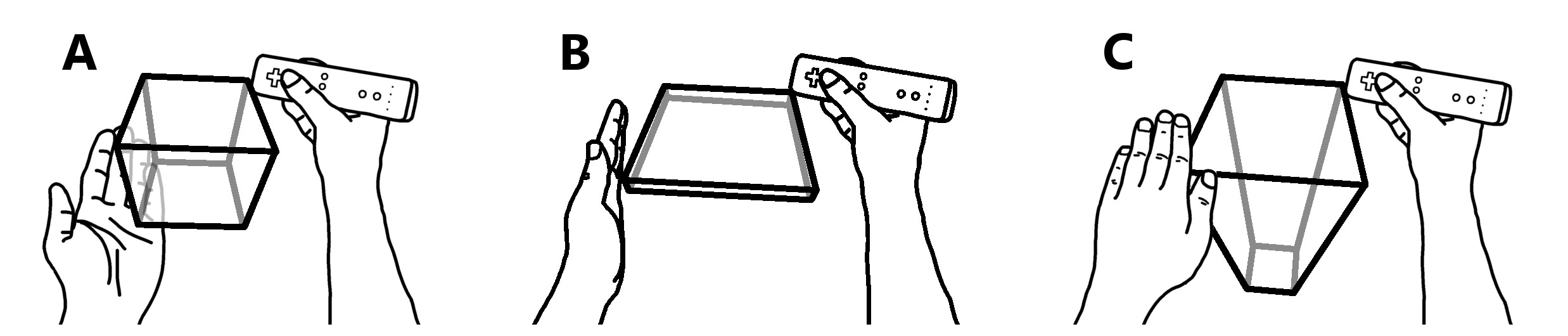}
  \caption{Three different bi-manual gestures to create a primitive: (A) non-dominant hand facing upwards, the primitive is created with uniform scale, (B) non-dominant hand facing the dominant hand, the primitive edges are located in each user's hand, to define size and height, (C) non-dominant hand facing down, the primitive bottom plane is snapped to the ground plane and the hands define the height and size.}
  \label{fig:gestures}
\end{figure*}

\begin{figure}[b!]
  \centering
  \includegraphics[width=\linewidth]{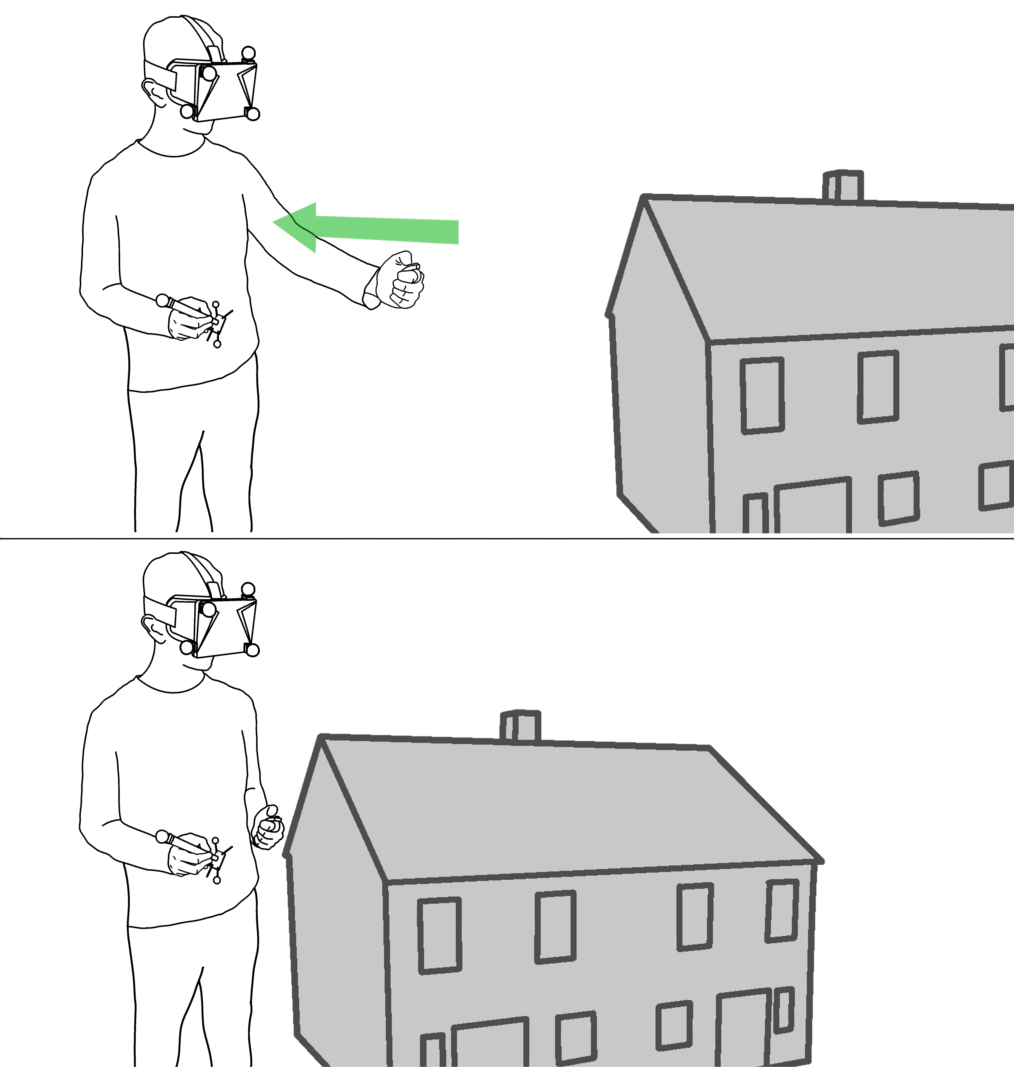}
  \caption{Pan gesture using the user's non-dominant hand.}
  \label{fig:world_pan}
\end{figure}

The world option activates the possibility of manipulating the virtual world. The user can manipulate the world by using the buttons on their non-dominant hand. 
To scale the world, the user presses the button and approximates the hands to lower the scale or moves away from the hands to increase it. The user can also reset the scale to the initial size by pressing the respective button in the menu.
While pressing the other button, the user controls the virtual world with a pan gesture. For example, by pressing the button at arm's length and making the gesture of pulling, like he is pulling the world to him, this gesture is depicted in Figure~\ref{fig:world_pan}, and it is an adaptation of the grabbing the air technique~\cite{Bowman:2004,Stoakley:1995}. The same behavior happens in any direction in the three different axes.

Since we allow the user to move the virtual world, he can lose the notion of the physical tracker limits. To keep the user aware of his physical movement limits, we always render a wired cube present like he is inside a glass box.

\begin{figure}[b!]
  \centering
  \includegraphics[width=\linewidth]{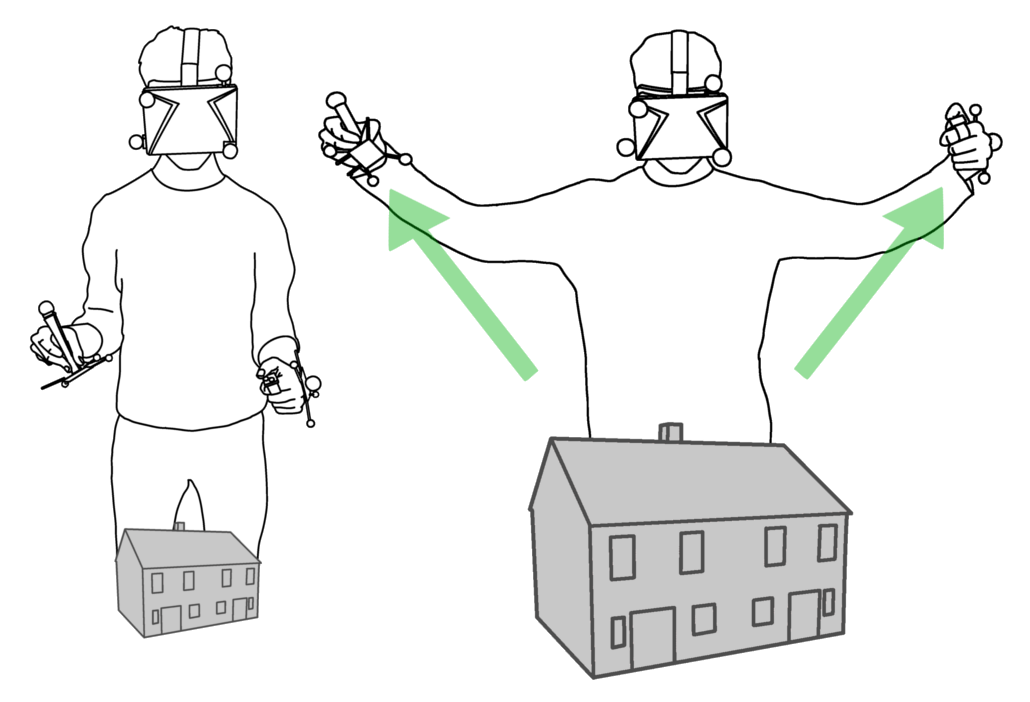}
  \caption{Scaling gesture using both hands.}
  \label{fig:setup_menu}
\end{figure}

\subsection{Create and manipulate}

The crate menu option allows the user to create one from a set of primitives. The user selects the primitive and uses bi-manual gestures to rapidly and naturally define the primitive's volume.
They can dynamically change the volume creation method using the non-dominant hand orientation. If the palm of their non-dominant hand faces the dominant hand, the volume of the primitives will be defined between both hands, like the opposite edges were located in each hand.
This way, a person can freely define the primitive volume with a bi-manual gesture (Figure~\ref{fig:gestures}.B).
However, if the non-dominant hand faces upwards, the primitive will always maintain a uniform scale, and the person can use the distance between both hands to define the size (Figure~\ref{fig:gestures}.A). In this case, the primitive changes color to gold for additional visual feedback.
If the non-dominant hand faces downwards, the primitive will be created from the ground plane, and the user defines the size and height using both hands' positions (Figure~\ref{fig:gestures}.C).
While the user defines the primitive, its material is semi-transparent, so they can see through it and help position the object where they want to. When they finalize the primitive, it turns into a solid material.
Before using the bi-manual gestures approach to dynamically switch between the primitive creation methods, we tried using the left-hand buttons to switch between the modes. However, users had more difficulties mapping the buttons to the desired creation method than with the gestures, which tended to be more immediate.

By changing the pen mode to selection, the user can manipulate every primitive he created and any pre-existing objects in the scene in 6-DOF. To this end, the user moves the Wiimote inside the primitive. When there is a collision between the object and the user's dominant hand pen virtual representation, the object's color slightly changes to give additional visual feedback to the user. The user needs to press the Wiimote button and select the object, showing a highlight shading around it. This procedure can be applied to multiple objects at the same time.
The user can then freely manipulate the selected objects by pressing the other Wiimote button. This technique is based on the Virtual Hand technique~\cite{Bowman:2004}.
We also allow the user to snap the object he is manipulating to any other existing one by facing the sides of the object to each other.
While the user is manipulating an object, the non-dominant hand menu turns into a garbage bin. If the user releases an object inside the bin, it is destroyed. This method lets the user remove any unwanted object from the virtual scene.

\section{Preliminary Evaluation}

We prepared a qualitative study with inexperienced users to evaluate our system's potential. Our goal was to understand their receptivity to our system.
In this evaluation, we wanted to verify whether our system's features were easy and useful for users and whether they could really help them complete some tasks.

\subsection{Apparatus}

For this evaluation, we used our setup located in our lab, as mentioned above. Thus, it guarantees a calm and controlled environment.
Eleven subjects, two males and two females, participated in our experiments. Since we wanted to evaluate our system with potential users, we gathered a group of six architects from a PhD course and five engineers from both Masters and PhD courses. Participants ranged from 25 to 60 years old (where 82\% were between 25 and 34 years old). None of the selected subjects had contact with our prototype before; only one had previously experienced head-mounted displays with no position tracking.

\subsection{Methodology}

Our evaluation included three different scenarios where users could sketch. The first was an enclosed virtual room, encouraging the user to move around the physical space, while the second and third were open virtual spaces requiring navigation features.

\begin{figure}[t!]
  \centering
  \includegraphics[width=\linewidth]{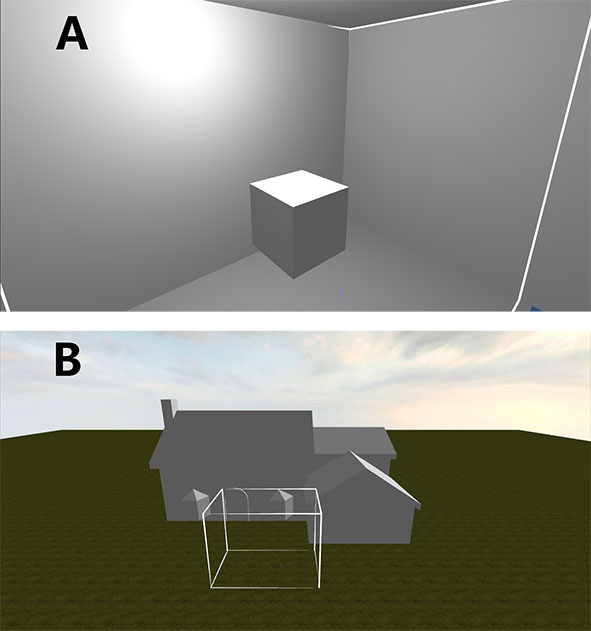}
  \caption{Scenarios used in our evaluation. One consists of a closed room with a floating cube (A), and the other is an open field with a model of a house (B)}
  \label{fig:scenarios}
\end{figure}

The first scenario consisted of a closed room, where the surrounding walls represent the limits of the physical room. This scenario also contains a floating cube in the middle, which the user can freely go through, Figure~\ref{fig:scenarios}.
In this scenario, we explained how the system works without explaining the God mode features and let users freely sketch and get used to the system for fifteen minutes. After this time, we asked users to draw in the floating cube sides, adding doors, windows, and a roof. During this trial, we encouraged users to try using the auxiliary plane tool and switch between pen and laser tools.

In the second scenario, subjects had an open space with a 3D model of a house without textures. Here, we explained the features of world manipulation and scale and how they work. We gave the user ten minutes to get used to it. After being ready, we asked to draw a door and two windows in an existing yellow wall of the house. After completing this task, we asked to create an extension to the house (a cube coming out of the wall) in a red wall on the other side of the house, using primitives. 

For the third task, in an empty scenario, we asked the users to be creative and create a small neighborhood using the features of our prototype, with at least five different objects representing different buildings.

After completing the tasks, subjects were asked to answer a questionnaire. Overall, each session in the experiment took around fifty minutes, although no time restrictions were imposed.

\subsection{Results}

The questionnaire users answered was composed of questions with six values Likert scale (1~-~strongly disagree, 6~-~strongly agree) to classify some aspects of the system and open-answer questions. From these surveys, we could collect interesting results.

None of the subjects showed relevant body fatigue, but when asked, they revealed that they felt more tired after using the system but did not feel it while using it. They indicated that this fatigue was especially caused by eyestrain due to the head-mounted display.
Since our prototype setup is wireless, we asked our subjects if they felt any delay between their movements and what they saw through the head-mounted display. 63\% of the subjects didn't feel any input lag; the others only reported minor momentary lag, probably related to the momentary loss of a rigid body by the tracking system.

\begin{figure}[t!]
  \centering
  \includegraphics[trim = 00mm 15mm 10mm 20mm, clip, width=\linewidth]{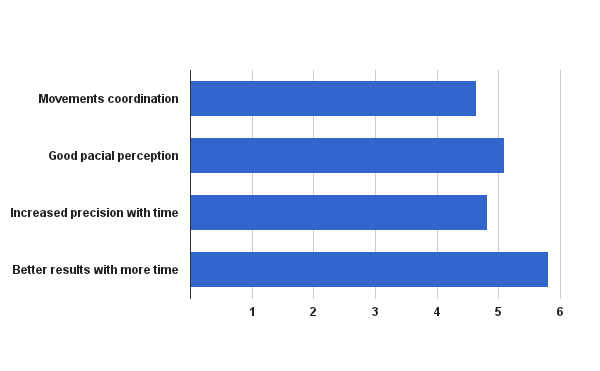}
  \caption{Mean values for perception and system acquaintance along time.}
  \label{fig:easyness_chart}
\end{figure}

Regarding the system acquaintance, subjects strongly agree that with more time, their results would certainly be better (Figure~\ref{fig:easyness_chart}).
From this chart, it is also possible to verify that subjects agreed that as long as they got more used to the system, the sketches became more precise. This fact was already evidenced in previous studies~\cite{Wiese:2010}.
Subjects also agreed that they had a good spatial perception of the virtual environment and could coordinate their movements with what they saw in the virtual world.

\begin{figure}[t!]
  \centering
  \includegraphics[trim = 00mm 15mm 36mm 20mm, clip,width=\linewidth]{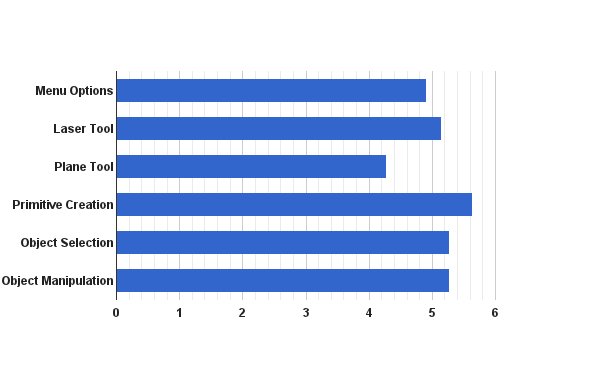}
  \caption{Mean times regarding tools easiness to use.}
  \label{fig:easytools_chart}
\end{figure}

\begin{figure}[b!]
  \centering
  \includegraphics[trim = 00mm 15mm 38mm 15mm, clip, width=\linewidth]{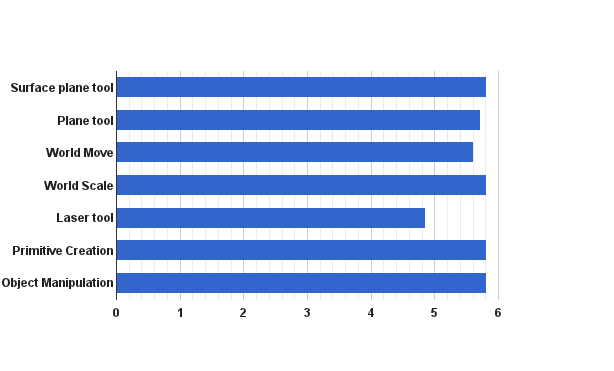}
  \caption{Mean times regarding tools usefulness.}
  \label{fig:tools_usefulness}
\end{figure}

From the graph of Figure~\ref{fig:easytools_chart}, we can notice that all subjects agreed the tools in our developed prototype are easy to use. This outcome results from the iterative process of developing our system.
It is also possible to notice that all the prototype's tools were considered useful, as shown in Figure~\ref{fig:tools_usefulness}. 

However, the laser tool was considered the least useful and easiest. The opposite happened with the plane tool, regarded as the least easy to use but the most useful tool.
One subject even said, during the experiment, that the laser tool does the same as the plane tool but with less accuracy since it can be used from far away.

\begin{figure}[t!]
  \centering
  \includegraphics[trim = 00mm 15mm 25mm 20mm, clip, width=\linewidth]{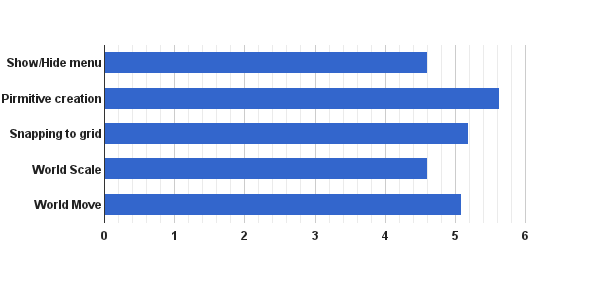}
  \caption{Mean times regarding gestures easiness to use.}
  \label{fig:gestures_easy}
\end{figure}

When asked about the bi-manual gestures implemented in our system, the subjects strongly agreed they were well mapped to the respective feature (mean time of 5,27 in 6 points).
From all the gestures in our prototype, the least liked by the subjects was the one to control the menu visibility and its relation to the pen being snapped to the proxy geometry plane tool when it was present (Figure~\ref{fig:gestures_easy}). Along with the gesture to control the menu visibility, the gesture to scale the world was also a least liked one. In this case, subjects complained that the virtual scale was happening too fast.
On the other hand, the gestures to change the primitive creation method were the most liked ones, and when asked if these gestures make the primitive creation task fast, they all strongly agreed (mean time 5,36 out of 6 points).

On the questionnaires, subjects highlighted our system's usefulness for 3D sketching and content creation, especially for 3D model design review. They also highlighted that with the fast primitive creation, they could easily add some volume detail to a model and sketch on it to add any necessary detail.


\subsection{Observations}

\begin{figure}[b!]
  \centering
  \includegraphics[width=\linewidth]{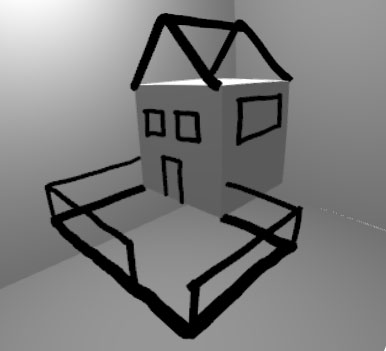}
  \caption{Example of a sketch made by one of the participants on the floating cube scenario.}
  \label{fig:house}
\end{figure}

\begin{figure}[t!]
  \centering
  \includegraphics[width=\linewidth,trim = 0mm 8mm 00mm 8mm, clip]{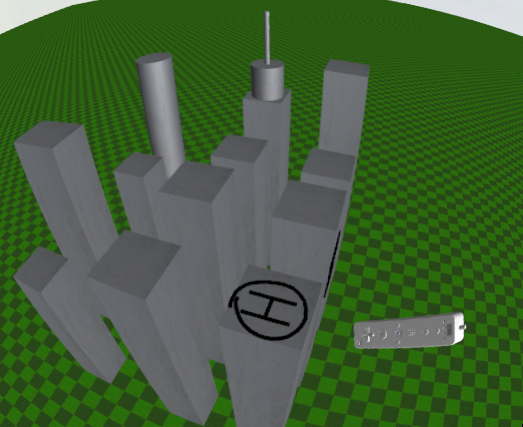}
  \caption{scenario created by a subject during our evaluation in task 3.}
  \label{fig:helli}
\end{figure}

From the subject's observations, we could gather some interesting observations.
Architects in our trials frequently mentioned the experience and said the system's potential was amazing.
One of them, who had experience in virtual reality systems, said that our system was the first real immersive experience he had ever had. This same subject was especially impressed by the lines' volume. Subjects frequently got engaged with our prototype so that they did more than they were asked to, as exemplified in Figure~\ref{fig:house}. In this example, besides drawing windows, doors, and roofs to the cube, a user added a fence using the plane tool several times. 
Another example is Figure~\ref{fig:helli}, where the subject sketched some high buildings, representing a neighborhood, for our third task.

The majority of the subjects had some depth perception difficulties when sketching. Usually, they were static while drawing in the intended space, and when they rotated or moved, they perceived that the line was not exactly in the desired place. This symptom was reduced after participants used the system for a while.

In general, subjects easily adapted to the auxiliary plane tool and frequently used it to aid in the desired tasks. One architect also said that it was the perfect sketching aid tool, and the plane's positioning snapping was useful because it allows one to sketch in parallel and perpendicular planes to existent sketches or surfaces. Another architect subject stated that the plane tool was more natural and easy to use than his CAD applications. 
However, only four subjects used the plane tool when sketching the door and windows in the house's yellow wall. The others preferred the laser tool far from the house, except for one subject who started using it and then decided to resort to the plane tool since it was more accurate.

We also noticed that, although participants could hide the menu easily, some tended to forget that they could do it; this was seen in our evaluation, where we noticed that this was the least liked gesture (Figure~\ref{fig:gestures_easy}. Subjects also tend to forget that they could use their body to move around the virtual world after we explain how they can use the ability to manipulate the virtual world position and scale. Two subjects even asked during the experiment how to rotate the world so they were facing the intended model surface instead of rotating their own body. This artifact might happen because we are used to controlling virtual environments while seated and looking at a static monitor.

Some users gave us suggestions on how to improve our system. One suggested that boolean operations between objects could be interesting and useful in opening more possibilities for content creation. Another suggestion was to have gestures to deform the primitives created.
They also suggested implementing more bi-manual gestures. One example was to face the hand palm forward while the user was creating a primitive object. This gesture would snap the primitive creation to the closest object in front of the user, similar to the facing-down gesture that snaps one's face to the ground plane.
Participants also suggested using the laser tool to select two or more points and then draw straight lines or curves between them, mimicking the pen's behavior when it is snapped to the auxiliary plane grid.

Other subjects suggested that it would be interesting to have a rubber tool to erase just some small spots of a drawn sketch instead of history. The same subjects stated the possibility of having a gesture to change the drawing color while sketching without relying on the menu to change the color. 
One user in our evaluation also suggested using a shake gesture to delete objects instead of our approach of releasing the objects inside the garbage bin.



\section{Conclusion and Future Work}

Sketching is an important part of several creative processes. In this work, we presented an immersive system for 3D air sketching. In our system, users can promptly materialize new ideas or review existing virtual models, sketching and creating primitives on and around them. We implemented several features to aid in this process, such as basic functionality to change stroke width and color, or more advanced ones, such as a proxy geometry plane tool for restricting sketches to two dimensions, distant lase-like sketching, primitive creation, and a navigate and scale mode to enable a vast virtual world. All these features in our prototype were developed using bi-manual gestures to make them easier and more natural for the users.

A preliminary evaluation of our system with six architects and five engineers showed that users enjoyed and found our immersive solution very useful. Users also thought of it as a very promising design review tool. The proxy geometry plane, primitive creation, and the ability to manipulate and scale the environment were very helpful. The subjects praised using bi-manual gestures to enhance the interactions, especially the ones used to create primitives differently.

We plan to add further capabilities to our prototype and perform more iterations on existing features. To increase its functionality and fulfill our vision to get closer to a world builder application, 
we intend to explore another type of geometrical content creation, such as planar or volumetric shapes, using 2D form recognition from the freehand sketches and combine this with object surface extrusion, similar to what has been done in Mockup Builder~\cite{DeAraujo:2012,Araujo2012-zp}. Also, adding a 3D object retrieval system inspired on~\cite{Fonseca2011-ma} will allow creating richer virtual environments. We feel that the possibility of using bi-manual gestures to deform existing primitives or to implement boolean operations between primitives could also be an interesting addition to our prototype. 

Furthermore, we plan to extend the concepts of proxy geometry and simple sketch-based construction rules to explore richer interaction and modeling scenarios. For example, while using the geometry plane proxy, explore other bi-manual gestures to allow curve definitions and not only lines. Still, regarding the auxiliary plane, we also think it would be useful to provide some gestures to allow the simultaneous placement of several planes to aid in more complex spatial drawings. 
We also want to explore the concept of object groups and develop bi-manual gestures to allow fast replication and dynamically change group copies.

We want to fully extend the physical interaction space to utilise our wireless VR setup. Combining these with augmented reality can also yield interesting results. We also think we can eliminate the virtual menu and develop an interface relying only on bi-manual gestures. It would also be interesting to evaluate both interaction techniques.



\bibliographystyle{acmsiggraph}
\bibliography{bibliography}
\end{document}